# Top Dielectric Induced Ambipolarity in an n-channel dual-gated Organic Field Effect Transistor

*Kaushik Bairagi [a,*], Elisabetta Zuccatti [a], Francesco Calavalle [a], Sara Catalano [a], Subir Parui [a †], Roger Llopis [a], Frank Ortmann [b], Fèlix Casanova [a,c], Luis E. Hueso [a,c]\**

[a] CIC nanoGUNE, 20018 San Sebastian, Spain

[b] Center for Advancing Electronics Dresden and Dresden Center for Computational Materials Science, Technische Universität Dresden, 01062 Dresden, Germany

[c] IKERBASQUE, Basque Foundation for Science, 48013 Bilbao, Spain

† Present address: IMEC and K. U. Leuven, Leuven 3001, Belgium

*E-mail: k.bairagi@nanogune.eu; l.hueso@nanogune.eu

**Abstract:**

The realization of both p-type and n-type operations in a single organic field effect transistor (OFET) is critical for simplifying the design of complex organic circuits. Typically, only p-type or n-type operation is realized in an OFET, while the respective counterpart is either suppressed by charge trapping or limited by the injection barrier with the electrodes. Here we show that only the presence of a top dielectric turns an n-type polymer semiconductor (N2200, Polyera ActiveInk™) into an ambipolar one, as detected from both bottom and top gated OFET operation. The effect is independent of the channel thickness and the top dielectric combinations. Variable temperature transfer characteristics show that both the electrons and holes can be equally transported through the bulk of the polymer semiconductor.

**Introduction**

Organic semiconductors (OSCs) are attractive for large area electronic devices over conventional inorganic semiconductors as they provide a better degree of freedom for tuning their mechanical, electrical and optoelectronic properties[1–8]. An interesting choice in the large family of OSCs is π-conjugated semiconducting polymers, since they are in general solution-processable, low cost, light weight and can be deposited on a variety of substrates and electrodes[9–14]. In most cases the semiconducting polymers are electronically intrinsic in nature and are potentially able to transport both electrons and holes [6,15,16], a capability which could be very useful for developing complex electronic devices with a single active material [17–20]. However, due to the presence of extrinsic charge trapping states, these materials are usually able to transport only one type of charge carriers [21,22]. Under certain conditions, these charge trapping states can be removed and the transport of both electrons and holes is possible through the semiconductor [22].

In organic field-effect transistors (OFETs), the routes to obtain ambipolar transport are commonly three-fold. The first case involves contact engineering so that both types of carriers can be efficiently injected [23]. The second case resorts to either blending p-type and n-type of molecules or to an ambipolar polymer synthesized by co-polymerization of individual donor and acceptor moieties [24–26]. The third case relies on specific engineering at the semiconductor/dielectric interface [10,15,16]. One such example of the last case is poly{[N,N′-bis(2-octyldodecyl)-naphthalene-1,4,5,8-bis(dicarboximide)-2,6-diyl]-alt-5,5′-(2,2′-bithiophene)} P(NDI2OD-T2) (commonly known as N2200, Polyera ActiveInk™, a highly electron transporting polymer) top-gated OFET in which a weak ambipolarity was observed using various polymer top gate dielectrics such as poly(methyl methacrylate) (PMMA) with $k$ = 3.6, Cyclized Transparent Optical

Polymer (CYTOP) with $k = 2.1$ and polystyrene (PS) with $k = 2.6$ [10]. In another experimental example, with the high-$k$ ferroelectric polymer (with $k = 10.8$ at 100 $kHz$) poly[(vinylidenefluoride-co-trifluoroethylene] (P(VDF-TrFE)) as a top gate, the hole mobility in N2200 was remarkably enhanced due to the change in dielectric polarizability at the semiconductor/dielectric interface[16]. However, there is no direct experimental evidence whether the dielectrics employed in the transistors actually change the bulk transport properties of N2200, turning it into an ambipolar material.

**Results and discussions**

Here we present the transistor operations of N2200 polymer in a lateral staggered geometry operating both in bottom-gate-bottom-contact (BGBC) and top-gate-bottom-contact (TGBC) configurations with which the effect of various top gate dielectrics on the N2200 channel can be studied. Figure 1b shows the schematic of such a dual-gated OFET with a 300 nm $SiO_2$ on $Si^{n++}$ as a global back gate and patterned Ti/Au (5 nm/36 nm) electrodes as source (S) and drain (D), a N2200 layer (36 nm) acts as the channel, different polymers as dielectric for the top gate and Al (25 nm) is the top gate contact. The dielectric polymers for the top gate were PMMA, bi-layer PMMA+CYTOP, bi-layer CYTOP+PMMA and CYTOP.

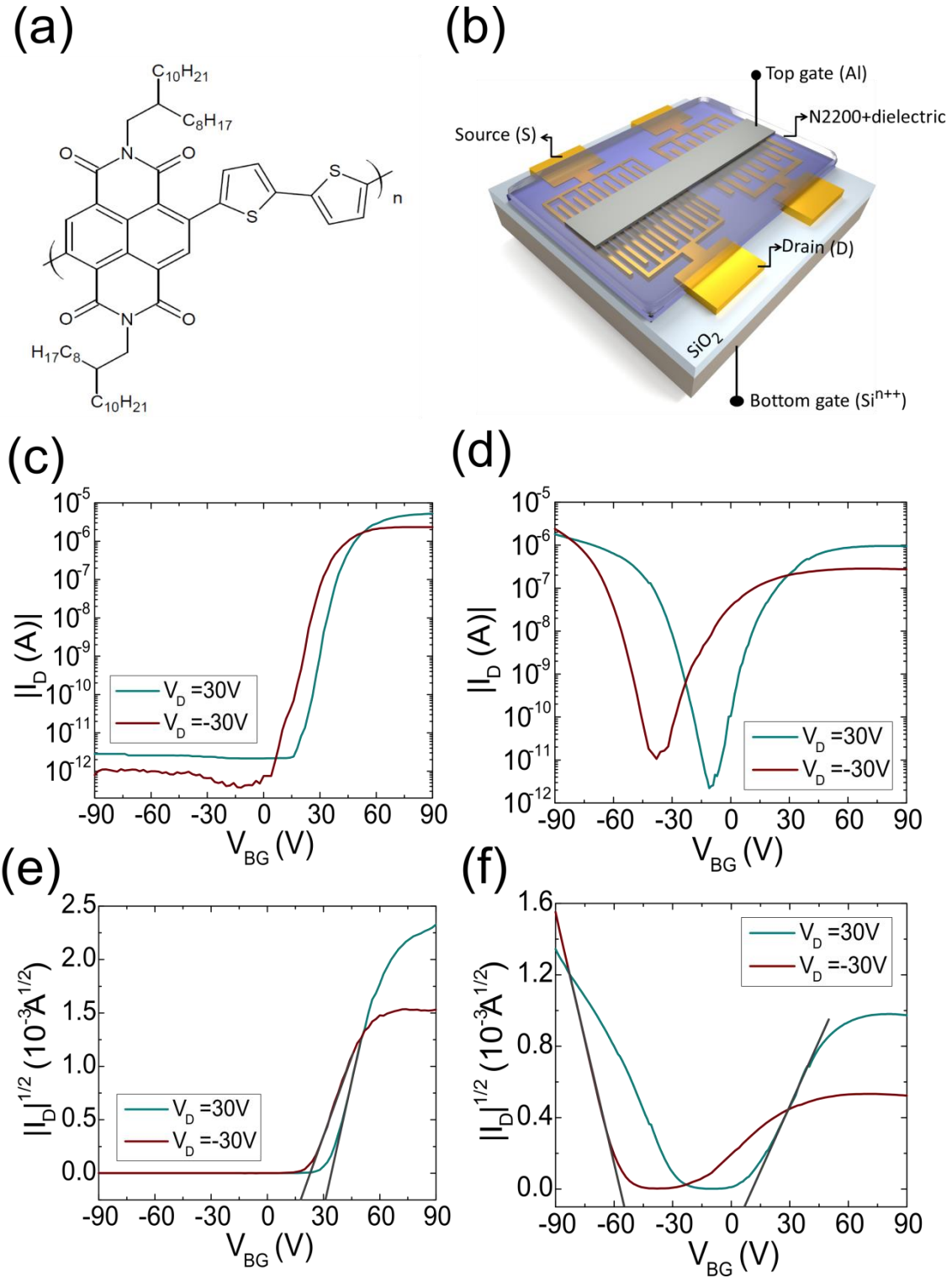

Figure 1. (a) Schematic diagram of the molecular structure of N2200. (b) Schematic presentation of the dual gate transistor architecture. (c) Transfer characteristics of the transistor (L=10 μm, W = 10 mm) measured with respect to the bottom gate (300 nm SiO$_2$ gate dielectric) with a N2200 (36 nm) channel only. (d) Transfer characteristics

measured with respect to the bottom gate with a PMMA top dielectric. (e) Square root of $I_D$ vs. $V_{BG}$ (from (c)). (f) Square root of $I_D$ vs. $V_{BG}$ (from (d)). The field effect mobilities in the saturation regime ($V_G - V_T < V_D$ where $V_G$: gate bias, $V_T$: threshold voltage and $V_D$: drain bias ) calculated from the slope of the curves (black lines) are $0.1\times10^{-3}$ cm$^2$V$^{-1}$s$^{-1}$ for electrons (from (e) without top PMMA) while $0.1\times10^{-3}$ cm$^2$V$^{-1}$s$^{-1}$ for electrons and $0.3\times10^{-3}$ cm$^2$V$^{-1}$s$^{-1}$ for holes (from (f) with top PMMA).

Figure 1c shows a typical n-type transfer curve of a N2200 OFET in BGBC mode with an ON-OFF ratio > $10^6$. The transistor (without top dielectric) does not show any sign of p-type behavior in any polarities of the drain bias ($V_D = +30V$ and $V_D = -30V$). On the contrary, the very same transistor with the presence of a top PMMA dielectric layer shows an ambipolar transfer characteristic in BGBC mode (Figure 1d) with an ON-OFF ratio $\approx 10^6$ for both p-channel ($V_D = -30V$) and n-channel ($V_D = +30V$) operations. In case of p-type operation, the turn on voltage ($V_{on}$) has a shift towards more negative gate bias which is consistent with previous studies [16,27]. The field-effect electron mobility at saturation for the n-channel operation without PMMA is $0.1\times10^{-3}$ cm$^2$V$^{-1}$s$^{-1}$ (Figure 1e). In the case of ambipolar operation with PMMA, the electron mobility for the n-channel operation (at $V_D = +30V$) is $0.1\times10^{-3}$ cm$^2$V$^{-1}$s$^{-1}$ while the hole mobility for the p-channel operation (at $V_D = -30V$) is $0.3\times10^{-3}$ cm$^2$V$^{-1}$s$^{-1}$ (Figure 1f). Thus, in the BGBC operation, there is an enhancement of the hole transport in the channel without any degradation of the electron transport. This behavior suggests that there is a substantial change in the bulk of the polymer rather than an interfacial effect at the semiconductor/dielectric interface.

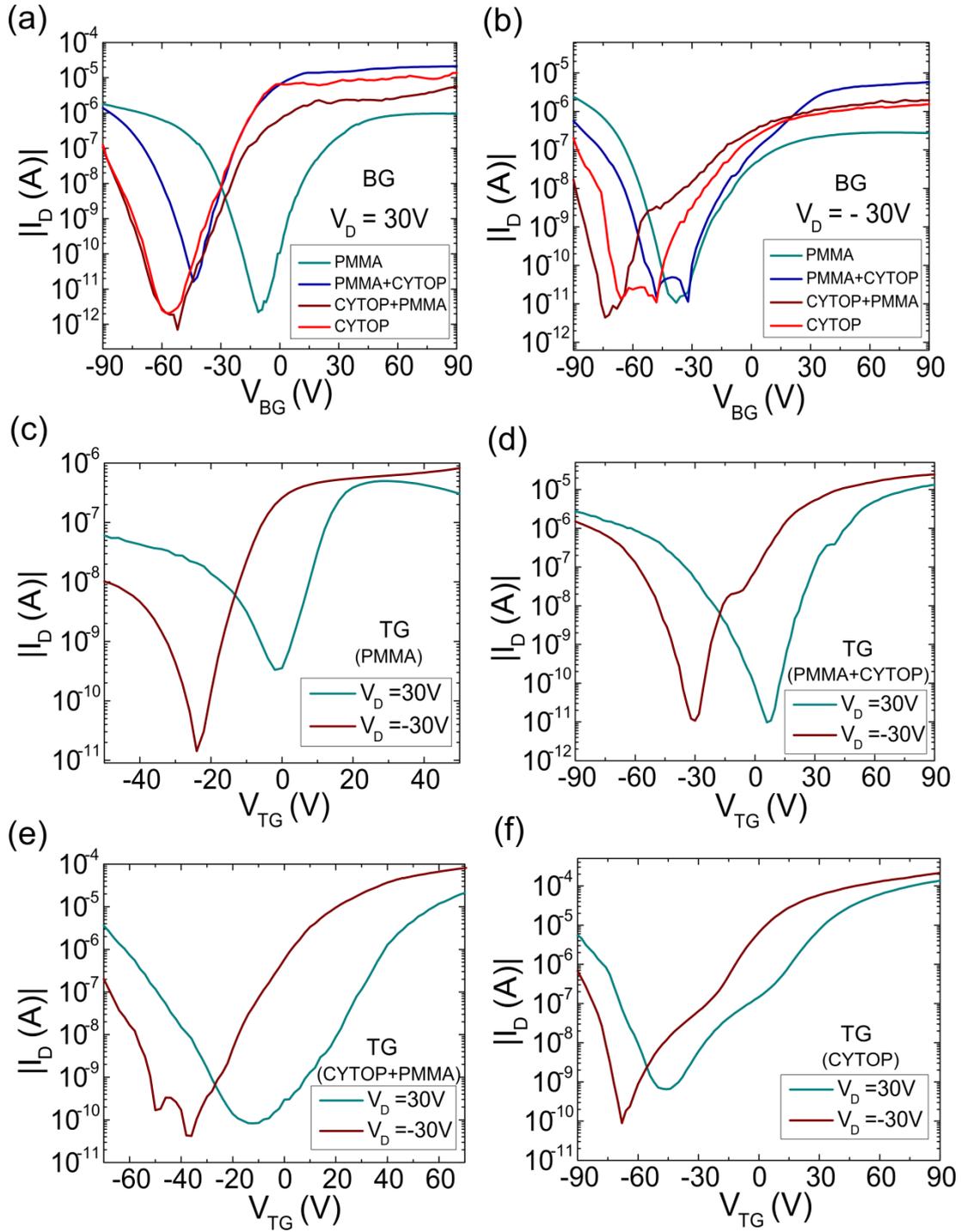

Figure 2. Transfer characteristics of N2200 (36 nm) transistor (probed with respect to the bottom gate having different top gate dielectrics with a drain bias $V_D = +30V$ (a), with $V_D = -30V$ (b). Transfer characteristics of the same transistors measured with respect to the top gate with different top dielectrics, (c) PMMA, (d) PMMA+CYTOP (e) CYTOP+PMMA and (f) CYTOP.

To understand the ambipolarity in N2200 in the presence of a top dielectric layer we have studied the OFET operation with another dielectric, CYTOP, and its bi-layer combination with PMMA – PMMA+CYTOP and CYTOP+PMMA. Figure 2a and 2b present the ambipolar transfer characteristics of the OFETs with all the dielectric combinations on N2200 channel at $V_D = +30V$ and at $V_D = -30V$ in BGBC mode. In all cases the ON-OFF ratio reaches $\approx 10^6$ with a very low off state current of the order of few pA (current density $\approx 10$ µA.m$^{-2}$).

To further support our previous hypothesis that the bulk of N2200 is modified, we have also studied the TGBC operation of the same devices gated with the top Al contact (Figure 1b). The transfer characteristics of all the transistors with top dielectrics PMMA, PMMA+CYTOP, CYTOP+PMMA, CYTOP are shown in figure 2(b)-2(f), respectively. Typical ambipolar transfer characteristics in all cases further confirm that the N2200 channel became throughout ambipolar in the presence of the top dielectric. The shift of the turn on voltages for p-channel with respect to n-channel towards more negative values is observed here as well. The field effect mobilities (for both electrons and holes) and the turn on voltages for both p-channel and n-channel in both BGBC and TGBC mode of operations are summarized in table 1a and table 1b, respectively. The bi-layer combination of the dielectrics is more favorable for both TG and BG operation of the devices with higher ON-OFF ratio and better charge carrier mobilities.

**Table 1a.** Mobilities and turn on voltages for N2200 BGBC OFET

| Top dielectrics | p-channel ($V_D = -30V$) | | n-channel ($V_D = 30V$) | |
|---|---|---|---|---|
| | $\mu_{h,BG}$ (× $10^{-3}$ cm$^2$ V$^{-1}$ s$^{-1}$) | $V_{on}$ (V) | $\mu_{e,BG}$ (× $10^{-3}$ cm$^2$ V$^{-1}$ s$^{-1}$) | $V_{on}$ (V) |
| NO DIELECTRIC | - | - | 0.1 | 15 |
| PMMA | 0.3 | -38 | 0.1 | -10 |
| PMMA+CYTOP | 0.6 | -48 | 2.1 | -43 |
| CYTOP+PMMA | 0.3 | -65 | 2.0 | -52 |
| CYTOP | 0.1 | -74 | 0.2 | -56 |

**Table 1b.** Mobilities and turn on voltages for N2200 TGBC OFET

| Top gate dielectrics | p-channel ($V_D = -30V$) | | n-channel ($V_D = 30V$) | |
|---|---|---|---|---|
| | $\mu_{h,TG}$ (× $10^{-3}$ cm$^2$ V$^{-1}$ s$^{-1}$) | $V_{on}$ (V) | $\mu_{e,TG}$ (× $10^{-3}$ cm$^2$ V$^{-1}$ s$^{-1}$) | $V_{on}$ (V) |
| PMMA | 0.4 | -24 | 13.1 | 0 |
| PMMA+CYTOP | 2.2 | -31 | 9.4 | 7 |
| CYTOP+PMMA | 10.5 | -37 | 11.1 | 10 |
| CYTOP | 6.8 | -68 | 19.9 | -45 |

Devices with two other different polymer channel thicknesses (55 nm and 115 nm) show similar performance to that of 36 nm presented above for all dielectric combinations (see supplementary information for details). The transport mechanism for both electrons and holes seems to be independent of the range of the channel thicknesses adopted in this study, which shows again that there is a change in the bulk of the N2200 polymer itself in the presence of the top dielectric.

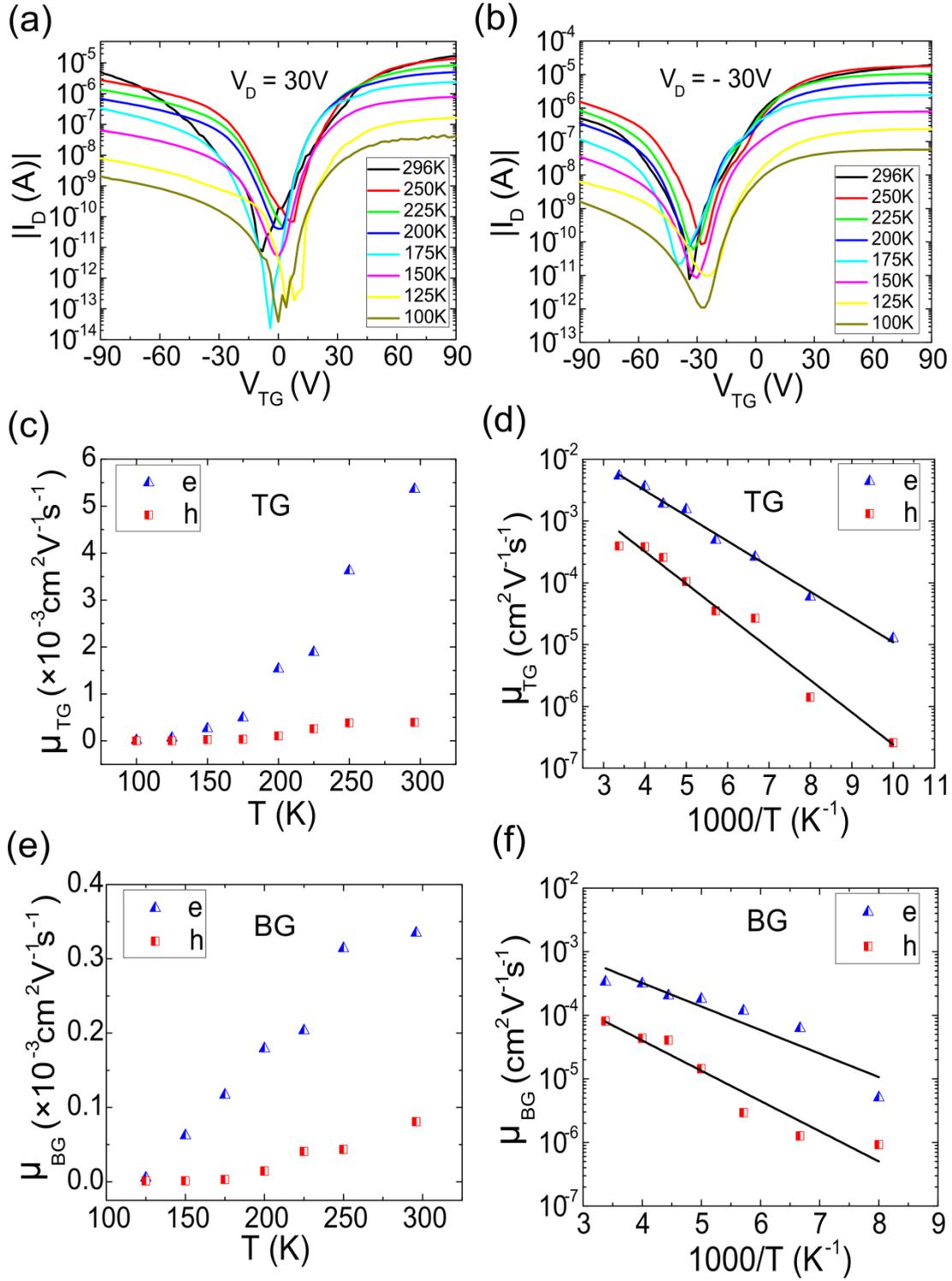

Figure 3. Temperature dependence of the transfer characteristics of N2200 (55 nm) transistor probed with respect to top gate (PMMA+CYTOP as the dielectric) with a drain bias $V_D = + 30V$ (a), with $V_D = -30V$ (b). Temperature dependence of hole and electron mobilities while the transistor is gated from top (c) and from bottom (e).

Semilogarithmic plot of mobility (symbols) of electrons and holes as a function of 1000/T and the corresponding Arrhenius fit (lines) while the transistor gated from top (d) and from bottom (f).

We present now the temperature dependent transport characteristics for the top and bottom gated transistors. We focus in the specific case of a 55 nm polymer channel thickness since the transport mechanism is similar in other samples. Figure 3a and 3b plot the temperature dependence of the transfer characteristics for a PMMA+CYTOP top-gated OFET for n-channel ($V_D$ = 30V) and p-channel ($V_D$ = -30V) operations respectively. The previously observed difference between the n-channel and p-channel turn on voltages is almost constant with temperature, indicating that the dielectric-semiconductor interface does not have deep states in which carriers can be trapped as the temperature is reduced [28]. The mobility of both electrons and holes follow a thermally activated behavior with an Arrhenius-like dependence (Figures 3c and 3d) [26,29]. Similar behavior is observed for the bottom-gated OFET (figure 3e and 3f), while the activation energies for electrons and holes for both the gating conditions are summarized in table 2. Both the similar values of mobility and its dependence with temperature strengthen the fact that the polymer itself becomes ambipolar in the presence of the top dielectric and that both types of carriers are transported through the channel.

Table 2. Activation energy ($E_A$) of the mobility for electrons and holes extracted from the slope of the curves in figure 3f and 3d for both BG and TG gating conditions.

| Gating condition | Electrons $E_A$ (meV) | Holes $E_A$ (meV) |
|---|---|---|
| BG | 31.9 ± 5.1 | 40.9 ± 5.1 |
| TG | 35.3 ± 1.1 | 44.8 ± 3.0 |

We observe that the mobility activation energy ($E_A$) for electrons and for holes is approximately similar in both gating conditions. The obtained values are relatively small and are comparable to other polymeric semiconductors [30–32] which indicates again that there is a low degree of interfacial dipolar disorder at the semiconductor/dielectric interface (see supplementary information for control experiments, including the role of the dielectric solvent).

We now focus on the possible reason for bulk hole and electron transport in a single N2200 transistor. In the rigid band approximation, the Fermi energy ($E_F$) of the Au contacts (5.0 eV) is almost at the middle of the band gap of N2200 whose highest occupied molecular orbital (HOMO) is at 5.6 eV and lowest unoccupied molecular orbital (LUMO) is at 4.0 eV [10], favoring the injection of both electrons and holes into the semiconductor. The transport mode of charge carriers in N2200 can be understood based on the reported surface analysis studies [27,33]. In one case, a face-on packing of the polymer on $Si^{n++}/SiO_2$ is observed with the molecular layers π-stacked in the out-of-plane direction [27], while in another study the growth of N2200 on Si substrate shows an edge-on orientation at the surface of the film while a face-on packing is observed in the bulk[33]. Although the face-on type molecular packing is not favorable for in-plane charge transport, the charge carriers can hop from one layer to the next one since they are coupled to each other via π-stacking. This electronic transport would lead to an out-of-plane mobility, evident from the measurements in vertical diode structures [15,27]. This suggests that in this case more than one molecular layer is taking part in the charge transport and that the transport mode is 3D (bulk) rather than a 2D (surface), as commonly observed in other edge-on stacked polymers such as P3HT and PBTTT [27,34,35]. Indeed, the AFM images of N2200 on $Si^{n++}/SiO_2$ (see supplementary information) show very good inter grain connectivity with low roughness which is favorable for a bulk transport mode. However, the bulk transport of both types of carriers in the same device is generally obscured by the charge trapping effect. Now, the presence of a top dielectric on N2200 removes the charge trapping centers for holes favoring its transport through the bulk of the semiconductor and thus irrespective of the gating conditions the polymer exhibits ambipolarity.

**Conclusions**

In conclusion, we have shown that the presence of a top dielectric drastically changes the charge carrier transport properties of N2200, thereby exhibiting ambipolarity both in bottom and top gated lateral transistor geometries. The channel thickness does not play a significant role in the charge carrier transport and ambipolarity is induced in the bulk of the semiconductor due to the top dielectric. The temperature dependent mobility for both electrons and holes show that both types of carriers are transported via thermally activated charge hopping process through the semiconductor thereby strengthening the fact that the polymer turned into an ambipolar material. Our results demonstrate that ambipolarity can be induced in the bulk of an n-type polymer semiconductor.

**Experimental**

*Transistor fabrication:* $Si^{n++}/SiO_2$ (300 nm) wafer was used to fabricate individual 1×1 $cm^2$ chips where $Si^{n++}$ was used as the global bottom gate with $SiO_2$ (300 nm) as the bottom gate dielectric. The patterned Ti/Au (5 nm/ 36 nm) electrodes, used as source (S) and drain (D), were fabricated using conventional lift-off photolithography technique. The individual chips were cleaned with isopropanol and dried with $N_2$ blow before spin coating the polymer. N2200 was dissolved in anhydrous chloroform and several solutions of 4 mg/mL, 6 mg/mL and 8 mg/mL were prepared to deliver thicknesses of 36 nm, 55 nm and 115 nm respectively. The solutions were subsequently spin-coated on the pre-patterned (with S and D electrodes) on $Si^{n++}/SiO_2$ (300 nm) substrates. The samples were then baked at 100°C for 30 minutes in cleanroom environment. PMMA and CYTOP were used as top gate dielectric without further purifications. After PMMA spin-coating, the sample was baked at 180°C for 2 minutes and after CYTOP spin-coating, the sample was baked at 130°C for 1h 30 minutes. After the top gate dielectric deposition, the Al top gate contact (25 nm) was thermally evaporated in an ultrahigh vacuum chamber on the transistor array using a shadow-masking technique.

***Thin films and transistor characterization:*** The thicknesses of all the N2200 films and PMMA films (60 nm and 160 nm) grown by spin-coating on $Si^{n++}/SiO_2$ (300 nm) was determined by X-ray reflectivity (XRR) technique. The thickness of spin-coated CYTOP (630 nm) was determined by profilometry. The electrical characteristics of the transistors were measured using Keithley 4200-SCS semiconductor analyzer in a variable temperature Lakeshore probe station under high vacuum.

**Conflicts of interest**
There are no conflicts to declare.

**Acknowledgements**
This work is supported by the European Research Council (Grant 257654-SPINTROS), by the European Commission (H2020-MSCA-ITN-2017- 766025-QuESTech) and by the Spanish Ministry of Science under Project RTI2018-094861-B-100 and Maria de Maeztu Units of Excellence Program - MDM-2016-0618. The work at TU Dresden was supported by the Deutsche Forschungsgemeinschaft (grant OR 349/1-1).

**References:**

1. K. Yoshino, Y. Ohmori, A. Fujii and M. Ozaki, *Jpn. J. Appl. Phys.*, 2007, **46**, 5655–5673.

2. M. Cinchetti, V. A. Dediu and L. E. Hueso, *Nat. Mater.*, 2017, **16**, 507–515.

3. O. Ostroverkhova, *Chem. Rev.*, 2016, **116**, 13279–13412.

4. J. Feng, *APL Mater.*, 2014, **2**, 081801.

5. H. Hwang, Y. S. Shim, J. Choi, D. J. Lee, J. G. Kim, J. S. Lee, Y. W. Park and B.-K. Ju, *Nanoscale*, 2018, **10**, 19330–19337.

6. A. Atxabal, S. Braun, T. Arnold, X. Sun, S. Parui, X. Liu, C. Gozalvez, R. Llopis, A. Mateo-Alonso, F. Casanova, F. Ortmann, M. Fahlman and L. E. Hueso, *Adv. Mater.*, 2017, **29**, 1606901.

7. S. Parui, L. Pietrobon, D. Ciudad, S. Vélez, X. Sun, F. Casanova, P. Stoliar and L. E. Hueso, *Adv. Funct. Mater.*, 2015, **25**, 2972–2979.

8. S. Vélez, D. Ciudad, J. Island, M. Buscema, O. Txoperena, S. Parui, G. A. Steele, F. Casanova, H. S. J. van der Zant, A. Castellanos-Gomez and L. E. Hueso, *Nanoscale*, 2015, **7**, 15442–15449.

9. A. Troisi and A. Shaw, *J. Phys. Chem. Lett.*, 2016, **7**, 4689–4694.

10. H. Yan, Z. Chen, Y. Zheng, C. Newman, J. R. Quinn, F. Dötz, M. Kastler and A. Facchetti, *Nature*, 2009, **457**, 679–686.

11. M. Caironi, M. Bird, D. Fazzi, Z. Chen, R. Di Pietro, C. Newman, A. Facchetti and H. Sirringhaus, *Adv. Funct. Mater.*, 2011, **21**, 3371–3381.

12. S. Parui, M. Ribeiro, A. Atxabal, R. Llopis, F. Casanova and L. E. Hueso, *Nanoscale*, 2017, **9**, 10178–10185.

13. S. Parui, M. Ribeiro, A. Atxabal, K. Bairagi, E. Zuccatti, C. K. Safeer, R. Llopis, F. Casanova and L. E. Hueso, *Appl. Phys. Lett.*, , DOI:10.1063/1.5045497.

14. Y.-F. Lin, W. Li, S.-L. Li, Y. Xu, A. Aparecido-Ferreira, K. Komatsu, H. Sun, S. Nakaharai and K. Tsukagoshi, *Nanoscale*, 2014, **6**, 795–799.

15. G.-J. A. H. Wetzelaer, M. Kuik, Y. Olivier, V. Lemaur, J. Cornil, S. Fabiano, M. A. Loi and P. W. M. Blom, *Phys. Rev. B*, 2012, **86**, 165203.

16. K.-J. Baeg, D. Khim, S.-W. Jung, M. Kang, I.-K. You, D.-Y. Kim, A. Facchetti and Y.-Y. Noh, *Adv. Mater.*, 2012, **24**, 5433–5439.

17. J. Zaumseil, R. H. Friend and H. Sirringhaus, *Nat. Mater.*, 2006, **5**, 69–74.

18. B. Balan, C. Vijayakumar, A. Saeki, Y. Koizumi and S. Seki, *Macromolecules*,

2012, **45**, 2709–2719.

19  L. Bürgi, M. Turbiez, R. Pfeiffer, F. Bienewald, H.-J. Kirner and C. Winnewisser, *Adv. Mater.*, 2008, **20**, 2217–2224.

20  J. C. Bijleveld, A. P. Zoombelt, S. G. J. Mathijssen, M. M. Wienk, M. Turbiez, D. M. de Leeuw and R. A. J. Janssen, *J. Am. Chem. Soc.*, 2009, **131**, 16616–16617.

21  M. M. Mandoc, B. de Boer, G. Paasch and P. W. M. Blom, *Phys. Rev. B*, 2007, **75**, 193202.

22  J. Zaumseil and H. Sirringhaus, Chem. Rev., 2007, **107**, 4, 1296-1323.

23  T. Kanagasekaran, H. Shimotani, R. Shimizu, T. Hitosugi and K. Tanigaki, *Nat. Commun.*, 2017, **8**, 999.

24  R. P. Ortiz, H. Herrera, C. Seoane, J. L. Segura, A. Facchetti and T. J. Marks, *Chem. - A Eur. J.*, 2012, **18**, 532–543.

25  E. J. Meijer, D. M. de Leeuw, S. Setayesh, E. van Veenendaal, B.-H. Huisman, P. W. M. Blom, J. C. Hummelen, U. Scherf and T. M. Klapwijk, *Nat. Mater.*, 2003, **2**, 678–682.

26  H. Sirringhaus, M. Bird and N. Zhao, *Adv. Mater.*, 2010, **22**, 3893–3898.

27  J. Rivnay, M. F. Toney, Y. Zheng, I. V. Kauvar, Z. Chen, V. Wagner, A. Facchetti and A. Salleo, *Adv. Mater.*, 2010, **22**, 4359–4363.

28  J. A. Letizia, J. Rivnay, A. Facchetti, M. A. Ratner and T. J. Marks, *Adv. Funct. Mater.*, 2010, **20**, 50–58.

29  J. D. Yuen, R. Menon, N. E. Coates, E. B. Namdas, S. Cho, S. T. Hannahs, D. Moses and A. J. Heeger, *Nat. Mater.*, 2009, **8**, 572–575.

30  N. Zhao, Y.-Y. Noh, J.-F. Chang, M. Heeney, I. McCulloch and H. Sirringhaus, *Adv. Mater.*, 2009, **21**, 3759–3763.

31  J.-F. Chang, H. Sirringhaus, M. Giles, M. Heeney and I. McCulloch, *Phys. Rev. B*, 2007, **76**, 205204.

32  A. Salleo, T. W. Chen, A. R. Völkel, Y. Wu, P. Liu, B. S. Ong and R. A. Street, *Phys. Rev. B*, 2004, **70**, 115311.

33  T. Schuettfort, L. Thomsen and C. R. McNeill, *J. Am. Chem. Soc.*, 2013, **135**, 1092–1101.

34  T. Muck, J. Fritz and V. Wagner, *Appl. Phys. Lett.*, 2005, **86**, 232101.

35  G. Horowitz, M. E. Hajlaoui and R. Hajlaoui, *J. Appl. Phys.*, 2000, **87**, 4456–4463.